\documentclass[lettersize,journal]{IEEEtran}
\usepackage{amsmath,amsfonts}
\usepackage{algorithmic}
\usepackage{algorithm}
\usepackage{array}
\usepackage[caption=false,font=normalsize,labelfont=sf,textfont=sf]{subfig}
\usepackage{textcomp}
\usepackage{stfloats}
\usepackage{url}
\usepackage{verbatim}
\usepackage{graphicx}
\usepackage{cite}
\usepackage{orcidlink}
\begin{document}

\title{Understanding Distribution Structure on Calibrated Recommendation Systems}

\author{Diego Corr{\^e}a da Silva\,\orcidlink{0000-0001-7132-1977},
Denis Robson Dantas Boaventura\,\orcidlink{0000-0003-2771-1710}, 
Mayki dos Santos Oliveira\,\orcidlink{0009-0007-4680-2868}, 
Eduardo Ferreira da Silva\,\orcidlink{0000-0001-8911-5805}, 
Joel Machado Pires\,\orcidlink{0000-0002-8428-3516}, 
Frederico Araújo Durão\,\orcidlink{0000-0002-7766-6666},
~\IEEEmembership{Institute of Computing,~Federal University of Bahia, Brazil}
\thanks{Corresponding author is Diego Corr{\^e}a da Silva. diego.correa@ufba.br}
\thanks{Manuscript received April 19, 2021; revised August 16, 2021.}}

\markboth{IEEE TRANSACTIONS ..., VOL. xx, NO. xx, OCTOBER xx}%
{Shell \MakeLowercase{\textit{et al.}}: A Sample Article Using IEEEtran.cls for IEEE Journals}


\maketitle

\begin{abstract}
Traditional recommender systems aim to generate a recommendation list comprising the most relevant or similar items to the user's profile. These approaches can create recommendation lists that omit item genres from the less prominent areas of a user’s profile, thereby undermining the user’s experience. To solve this problem, the calibrated recommendation system provides a guarantee of including less representative areas in the recommended list. The calibrated context works with three distributions. The first is from the user's profile, the second is from the candidate items, and the last is from the recommendation list. These distributions are G-dimensional, where G is the total number of genres in the system. This high dimensionality requires a different evaluation method, considering that traditional recommenders operate in a one-dimensional data space. In this sense, we implement fifteen models that help to understand how these distributions are structured. We evaluate the users' patterns in three datasets from the movie domain. The results indicate that the models of outlier detection provide a better understanding of the structures. The calibrated system creates recommendation lists that act similarly to traditional recommendation lists, allowing users to change their groups of preferences to the same degree.
\end{abstract}



\section{Introduction}
\label{sec:intro}

Commonly, traditional recommender systems generate recommendations with \textit{miscalibration}~\cite{Lin:2020:Collaborative}. Miscalibration means that the recommendation lists do not follow the user preferences distribution, instead suggesting items from user's dominant area of interest. It creates an overspecialized recommendation list in which the items from the less dominant area are overwhelmed. This effect puts the user in a filter bubble or an echo chamber problem~\cite{Kamishima:2012:Bubble}. For instance, when a specific area dominates the recommended list, the user likely has few other options to interact with, aside from items within that dominant area. Then, the subsequent lists are recommended, with the dominant area becoming more overspecialized. In recent years, calibrated recommendation systems have attracted attention \cite{Kaya:2019:Intent, Abdollahpouri:2021:POP, Naghiaei:2022:Towards, Silva:2022:benchmark, Steck:2018:Calib, Silva:2025:Timeaware} from the recommender system community to overcome this issue. This type of system demonstrates the capacity to improve several objectives, such as diversity~\cite{Kaya:2019:Intent}, control of popularity bias~\cite{Abdollahpouri:2021:POP}, item coverage~\cite{Naghiaei:2022:Towards}, precision~\cite{Silva:2022:benchmark}, and the reduction of miscalibration~\cite{Steck:2018:Calib}.

To illustrate how calibrated recommendation works, consider a scenario: if a user's preferences distribution indicates 65\% for sci-fi and 35\% for adventure, an ideally calibrated recommendation list would aim to reflect these proportions. This preferences distribution presents the target proportions for the recommendation list. Thus, even with minor deviations, the recommendation list is designed to align with each user's actual preferences, giving items from less dominant categories more exposure.

State-of-the-art studies introduce many ways to extract the user preference distributions based on features such as genres~\cite{Steck:2018:Calib}, sub-profiles~\cite{Kaya:2019:Intent}, and item popularity~\cite{Abdollahpouri:2021:POP}. Among these, the use of genre-derived distributions is a prevalent approach and has been shown to improve multiple recommendation objectives~\cite{Steck:2018:Calib, Kaya:2019:Intent, Seymen:2021:Constraint, Silva:2021:Exploiting}. The users' distributions comprise patterns and behaviors through their structure and values. Some modeling structures can be employed to analyze these distributions, such as clustering, fuzzy logic, hierarchical approaches, graphs, or tree-based techniques. Each modeling structure is associated with specific algorithms and workflows. For example, clustering algorithms typically identify groups by assessing the distance or proximity between instances; that is, instances close to each other are assigned to the same cluster (group)~\cite{Arthur:2007:SODA}. Fuzzy algorithms determine the degree of pertinence of an instance to a particular cluster \cite{Dias:2019:Fuzzy}. Hierarchical algorithms connect and merge the groups of instances to form a tree-like structure of nested clusters \cite{Schubert:2021:LWDA}.

To the best of our knowledge, previous works have not examined the structure of user preference distribution or fully characterized their behavior. In \cite{Starychfojtu:2021:Recepies}, the genre information is used to construct recommendation lists, considering that the Fuzzy D'Hont algorithm can understand how to generate effective recommendations. However, these authors do not explain whether the distribution adheres to the fuzzy structure. Thus, this paper aims to investigate these distributions' structure and identify their fit patterns. In addition, the studies \cite{Abdollahpouri:2019:Unfairness, Naghiaei:2022:UnfairnessBook} categorize the users into three groups (niche, blockbuster, and diverse) based on the proportion of popular items in their preference profiles. This proportion is derived from a 1-dimensional boxplot analysis. Given that preference distributions in a calibrated recommendation context are often $G$-dimensional, where $G$ is the number of genres available in the system, the 1-dimensional boxplot analyses employed in prior state-of-the-art approaches are not directly applicable. This limitation motivates the current study to explore alternative techniques for identifying and understanding the user preference distribution structure.

User preference distributions can be derived from either original or binarized user scores, and their computation may involve a normalization step, depending on the method used. The work \cite{Silva:2022:benchmark} proposes a new function to extract distributions in the calibrated context. This function normalizes the distribution values after using the original score in the computation. However, this study does not investigate the effect of binarizing the scores, such as the approach applied in \cite{Steck:2018:Calib, Kaya:2019:Intent}. Therefore, in this paper, we compare these two types of distribution: a) those derived from original user scores and b) from binary scores.

Another important aspect is how calibrated systems are evaluated. In \cite{Steck:2018:Calib}, the miscalibration metric is used for evaluating calibrated recommender systems. In sequence, \cite{Silva:2021:Exploiting} introduces two metrics for this purpose, which consider the rank position and adapt to the divergence/similarity measure employed. Nevertheless, these studies do not consider distribution structures in their evaluation processes. To address this gap, this paper investigates methods for evaluating calibrated recommendation systems by incorporating their distribution structures and patterns.

Based on the preceding discussion, our investigation is guided by the following six research questions. First, we investigate whether connecting users based on their behaviors is possible, forming coherent groups. \textbf{RQ1:} How many groups are present in the user distribution patterns? Given that the recommendation process generates three distinct distributions, we examine whether users change behavioral groups across steps. \textbf{RQ2:} Do users switch groups when comparing the inherent distributions of candidate items and calibrated recommendations with their original preference distributions? We also explore the structure induced by the G-dimensional space generated by the system. \textbf{RQ3:} What kind of structure do user distributions follow? Since distribution values can be normalized or not, we examine whether the type of score affects the resulting structure. \textbf{RQ4:} Do different score types (e.g., binary vs. continuous) lead to different distribution structures? Following this, we examine whether such structural differences affect the performance metrics. \textbf{RQ5:} Which type of score results in better metric values? Finally, we investigate whether analyzing the distribution structure can serve as a basis for evaluating calibrated recommendation systems. \textbf{RQ6:} Can distribution structures be used to evaluate calibrated systems?

To answer these research questions, we investigate the underlying structures of user preference distributions within the context of calibrated recommendation systems. These systems generate three key distributions: (1) the user preference distribution, derived from their historical interactions; (2) the candidate item distribution, derived from the list produced by a traditional recommender algorithm (e.g., SVD); and (3) the calibrated recommendation list distribution, created by the calibrated system.

To identify the distribution type, we first analyze the users' preferences distributions to identify dominant structural patterns in the high-dimensional distributions, establishing them as the baseline structure. Next, we examine whether the candidate item distributions and the calibrated recommendation list distributions maintain consistency with this baseline or induce shifts in user preference groups. To achieve this, we employ 15 diverse algorithms—including clustering (e.g., K-Means), outlier detection (e.g., Isolation Forest), and hierarchical models (e.g., Agglomerative)—to label and compare the distributions. We evaluate structural adherence using metrics such as the Silhouette score for cluster cohesion and Jaccard similarity to quantify group changes.

This study contributes to a deeper understanding of how calibration structures operate and the development of calibrated recommender systems. The main contributions of this study are:

\begin{itemize}
    \item Identifying the underlying distribution structures used in calibrated recommender systems;
    \item Measuring the degree of conformity between recommendation lists and users' preference distributions;
    \item Analyzing how candidate items and recommended items shift according to users' preference distributions;
    \item Evaluating the effects of binary versus non-binary scores on distribution characteristics and metric outcomes;
    \item Estimating the number of user groups based on distributional patterns;
    \item Proposing new evaluation models that incorporate distribution structures into the assessment of calibrated systems.
\end{itemize}

The remainder of this article is organized as follows. Section \ref{sec:related_work} reviews prior research on calibrated recommender systems, fairness-aware recommendations, and distribution analysis in traditional recommendation systems. Section \ref{sec:related} formalizes the calibration method, detailing the distributions, trade-off optimization, and item selection algorithms. Section \ref{sec:back} explains the extraction of genre-based distributions from user preferences, candidate items, and recommendation lists, illustrated with numerical examples. Section \ref{sec:proposal} presents our structural analysis framework, introducing 15 algorithms (e.g., clustering, outlier detection) to model distribution patterns. Section \ref{sec:setup} describes the experimental setup, including datasets, preprocessing steps, and evaluation metrics. Section \ref{sec:results} answers the six research questions (RQ1–RQ6), analyzing group consistency, structural adherence, and the impact of score types (binary vs. original). Also, it synthesizes findings on user group dynamics, algorithmic performance, and implications for fairness and calibration. Finally, Section \ref{sec:final} concludes the paper by summarizing key contributions, discussing limitations, highlighting practical insights, and outlining future research directions (e.g., popularity-based distributions, cross-domain validation).

\section{Related Work}
\label{sec:related_work}
This section reviews the foundational literature for this work, addressing two main areas of research in recommender systems. First, we discuss the concept of fairness, exploring how algorithmic biases can arise and disproportionately impact different groups of users and items. Next, we jump into calibration techniques, a method often used to align recommendations with user preference distributions, which also serves as an approach to promote fairness.

\subsection{Fairness}
\label{sec:fairness}
Fairness in recommender systems has become a crucial area of study, driven by the growing awareness that algorithms can perpetuate or even amplify existing biases. Research in this field focuses on identifying, measuring, and mitigating the disparate treatment of users or the unfair exposure of items. The following studies explore various facets of this problem, from defining and quantifying unfairness to proposing methods for its reduction.

The research \cite{Abdollahpouri:2019:Unfairness} investigates unfairness in traditional recommender algorithms. In the study, the authors classify users into niche, diverse, and Blockbuster-focused groups. To form these groups, each user's profile is analyzed based on the proportion of popular items they consume. The niche group consists of users whose profiles are composed of more than half unpopular items (long-tail items). The Blockbuster-focused group includes the top 20\% of users with, on average, more than 85\% of popular items in their profiles. The ``diverse" type of users are all those not in the Niche or Blockbuster-focused groups. However, the analysis is limited to a one-dimensional problem, without debating issues based on multiple dimensions. Similarly, in \cite{Naghiaei:2022:UnfairnessBook}, the same group division is employed in the context of book recommendations, also keeping the analysis focused on a one-dimensional problem.

The work \cite{Yongsu:2024} proposes a unified framework to systematically examine the major difficulties and, most importantly, the stereotyping problem in recommender systems. Stereotyping is defined as the tendency of a system to overgeneralize a user's preferences based on their group membership (e.g., demographics), rather than reflecting their individual tastes. The framework decomposes the recommendation error into two distinct sources: Bias (systematic distortion of the system's mean predictions) and Variance (variability of individual predictions), associating system bias with the first component and stereotyping with the second. The analysis reveals that simpler algorithms tend to be more stereotypic, that minority groups and users with atypical tastes are disproportionately affected by these errors, and that an oversampling strategy can mitigate stereotyping, albeit with certain trade-offs.

In \cite{Abdollahpouri:2021}, the authors propose an aggregated diversification approach to meet strategic platform goals, such as promoting fairness for item providers through more equitable exposure. The proposal is a post-processing framework that calibrates recommendations to mitigate popularity bias, making the overall distribution of items across the system closer to a predefined target distribution. This target distribution, in practice, reflects the platform's fairness policy (e.g., giving more visibility to niche items). The authors explore two variations: one that calibrates recommendations solely toward this system fairness target, and a more sophisticated one that also considers the consumption profile and tolerance for diversity of each individual user. This allows for managing the trade-off between relevance and the fairness goal, resulting in less loss of accuracy and less individual miscalibration.

While the mentioned works address fairness by categorizing users based on one-dimensional metrics like popularity or by decomposing aggregated system errors, they do not investigate the fundamental and multidimensional structure of the users' preferences. In contrast, our proposal does not pre-define user groups based on a single bias, such as popularity. Instead, we apply clustering algorithms to discover the structures that naturally emerge from the users' preference distributions based on movie genres. By analyzing how these groups are formed and whether users transition between them during the recommendation process, our work offers a new perspective on fairness, focused on the intrinsic patterns of user taste, rather than being limited to a single, previously selected bias metric.

\subsection{Calibration}
\label{sec:calibration}
A calibrated recommendation is classified as a multi-objective recommender system (MORS)~\cite{Jannach:2022:MORS}. A MORS aims to create recommendation lists that go beyond simple relevance or similarity, potentially optimizing objectives such as diversity, novelty, serendipity, item coverage, personalization, and calibration. In particular, a calibrated recommender system uses the user's preference distribution as a target to generate the recommendation list, seeking to replicate this distribution as closely as possible. Several studies~\cite{Silva:2021:Exploiting, Yifan:2022:Survey, Evaggelia:2021:Overview, Deldjoo:2022:Survey, Yunqi:2022:Survey, Jannach:2022:MORS} point to calibration as a means of promoting fairness, classifying it within different categories of fairness.

Earlier Steck~\cite{Steck:2018:Calib} proposes a calibrated recommendation method to balance list relevance and calibration. To calibrate the list, the author uses an equation that extracts a distribution from the user's preferences, using it as a target. This distribution is derived from the scores the user gives to items and the genres of those items. Steck's method applies binarization to the scores to create a normalized distribution. To generate the recommendation list, it uses the scores predicted by the recommender algorithm and the item genres. The author finds that the calibrated recommendation reduces the discrepancy between the distributions but argues that the increase in calibration comes at the expense of precision. The study, however, does not explore other formulations for the preference distribution.

da Silva et al.~\cite{Silva:2021:Exploiting} develop two new metrics to evaluate calibrated recommendations, designed to measure the difference between distribution values considering rank position and aggregate this difference into a single value. The first, Mean Absolute Calibration Error (MACE), evaluates the reduction of the difference over absolute values. The second, Mean Rank MisCalibration (MRMC), measures miscalibration by considering rank evolution and the distance/similarity measure used. For both metrics, lower values indicate a better list. In contrast to \cite{Steck:2018:Calib}, the authors use the original score value given by the users, and the results show that calibrated recommendations can produce a positive increase in precision. However, the study does not evaluate whether the distributions have a specific structure or if the original score indeed produces better recommendations.

Souza and Manzato~\cite{Souza:2024} present a two-step calibration proposal to mitigate popularity bias and unfairness in recommender systems. The approach is applied in the post-processing stage, which allows its use with different recommendation algorithms. The methodology begins with the first stage (Popularity Calibration), where the initial list of candidate items is adjusted to reflect the user's preferences for different levels of popularity. The second stage (Genre Calibration) uses the intermediate list to adjust the distribution of genres according to the user's consumption profile, generating the final list. The method uses Kullback-Leibler (KL) divergence and introduces two trade-off weights to control the intensity of the popularity and genre calibrations, allowing for personalized adjustment.

In \cite{Lin:2024}, the authors investigate how the dynamic nature of user preferences affects the calibration of recommender systems. The work is based on the premise that standard calibration methods assume static preferences and measure calibration over the entire interaction history, which may include outdated data and distort the representation of users' current interests. To address this, the authors propose identifying the most relevant segment of a user's history to optimize calibration, searching for the ``time window" of interactions that best represents their current interests. To do this, the interaction history is divided into chronological segments, which are then iteratively combined to create different training samples. Each sample is used to train a model and evaluate its performance, measuring both miscalibration (using KL divergence) and accuracy.

The research \cite{Kleinberg:2024} address the challenge of calibrating recommendations in a scenario that considers user decaying attention, where items at the top of the list receive more attention. To solve this problem, they developed algorithms for two distinct models. In the Discrete Genre Model, where each item belongs to a single genre, the solution uses a novel bin-packing analysis, treating each genre as a "bin" where the attention weight of an item is allocated to match user preferences. In the Distributional Genre Model, where an item can be a mixture of several genres, the approach extends constrained submodular optimization, a technique ideal for problems with the "diminishing returns" property. To quantify calibration in both scenarios, the work introduces the formalism of overlap measures.

da Silva et al.~\cite{Silva:2025:Timeaware} propose new approaches for calibrating recommendations, focusing on two limitations of existing methods. The first is the assumption that user preferences are static, ignoring that they can change over time. The second is the way interests are calculated for items with multiple attributes (such as a movie with multiple genres), which can lead to an inaccurate representation of the user profile. To address this, the authors introduce two classes of techniques: temporal approaches, which assign greater weight to recent interactions or analyze preference phases, and an entropy-based approach (GLEB), which balances the importance of an attribute in a specific item (``local") with its importance across the entire user profile (``global").

da Silva and Jannach~\cite{Silva:2025:Survey} provide a comprehensive survey of the state-of-the-art in calibrated recommendations. They analyze each published paper individually, covering all works since 2018, as well as earlier studies that introduced related ideas prior to Steck’s foundational work~\cite{Steck:2018:Calib}. The survey highlights that no existing research explicitly explores the structure of the user preference distribution. Additionally, the authors outline promising future directions for advancing the field.

These studies have significantly advanced calibration techniques, introducing new metrics, optimizations, and considering factors such as temporality. However, a common limitation is that they evaluate the success of calibration by focusing on the final outcomes, without examining the structural properties of the distributions during the process. Fundamental works like those previously presented did not investigate whether these multidimensional distributions adhere to specific patterns. Our research directly addresses this gap by analyzing the structural characteristics of the three key distributions (preferences, candidates, and final list). We investigate whether they form coherent groups, the nature of these structures, and how they are impacted by different score formulations, thus providing a complementary and more in-depth evaluation than existing approaches.

\section{Calibration Method}
\label{sec:related}

Steck \cite{Steck:2018:Calib} starts the research on the calibrated recommendation system, proposing two distributions, $P(g|u)$ derived from the user's preferences and $Q(g|u)$ derived from the user's recommendation list, as described below:

\begin{equation}\label{eq:weighted_distribution_p}
P(g|u) = \frac{\sum_{i}^{I(u)}w_{u,i}\cdot p(g|i)}{\sum_{i}^{I(u)}w_{u,i}},
\end{equation}

\begin{equation}\label{eq:weighted_distribution_q}
Q(g|u) = \frac{\sum_{i}^{L(u)}w_{r(i)}\cdot p(g|i)}{\sum_{i}^{L(u)}w_{r(i)}},
\end{equation}

\begin{equation}\label{eq:weighted_distribution_tildeq}
\tilde{Q}(g|u) = (1-\alpha)\cdot Q(G|u) + \alpha \cdot P(c|u).
\end{equation}

\noindent where $I(u)$ represents the user $u$ preferences, $w_{u,i}$ is the score that $u$ gives to item $i$, $p(g|i)$ is the proportion of genre $g$ in $i$ (commonly $\frac{1}{|i\ total\ genres|}$), $L(u)$ represents the $u$ recommendation list, $w_{r(i)}$ is the predicted score by the recommender algorithm that maybe $u$ gives to $i$ and $\alpha$ is a value which normally is $0.01$.

Most studies on calibrated recommendations use these distributions based on genres \cite{Steck:2018:Calib, Silva:2021:Exploiting, Naghiaei:2022:Towards}. Other works, such as \cite{Kaya:2019:Intent} and \cite{Abdollahpouri:2021:POP}, instead of genres implement sub-profiles and popularity sequentially. Thus, we can infer that any class, category, label, genre, or tag can be used.

To measure how calibrated $Q(g|u)$ is with $P(g|u)$, Steck~\cite{Steck:2018:Calib} proposes to use the divergence measure called Kullback-Leibler. The work \cite{Silva:2022:benchmark} shows that to compute if $Q(g|u)$ is calibrated with $P(g|u)$, any similarity or divergence measure can be used. They implement 57 measures and indicate that choosing the right measure can double the precision. The authors conclude that the equation below achieves the best precision:

\begin{equation}\label{eq:vicisSymmetricChiSquareEmanon2}
C_{emanon2}(P, Q) = \sum_{g}^{G} \frac{(P(g|u) - Q(g|u))^2}{\min(P(g|u), Q(g|u))^2}.
\end{equation}

The calibrated recommendation system is a MORS~\cite{Jannach:2022:MORS}, which implies that the system creates a recommendation list with more than one objective. The main focus of the calibrated system is to provide calibration and relevance. So, the relevance function can be described in many equations. In \cite{Steck:2018:Calib}, the predicted scores of the user's recommendation list are summed. The work \cite{Seymen:2021:Constraint} conducts and infers changes in the Steck proposal. Abdollahpouri et al.~\cite{Abdollahpouri:2023:MinCost} propose to use a latent factor approach to find the list relevance. da Silva and Durao~\cite{Silva:2022:benchmark} implement the Normalized Discount Cumulative Gain (NDCG), which is defined as below:

\begin{equation}\label{eq:dcg}
DCG(L) = \sum_{i}^{L}\frac{2^{w_{r(i)}} - 1}{\log_{2}^{i + 1}};
\end{equation}

\begin{equation}\label{eq:idcg}
iDCG(L) = \sum_{i}^{L^{ideal}}\frac{2^{w_{r(i)}} - 1}{\log_{2}^{i + 1}};
\end{equation}

\begin{equation}\label{eq:ndcg}
S_{ndcg}(L) = NDCG = \frac{DCG}{iDCG}.
\end{equation}

The $C_{emanon2}(P, Q)$ measures the calibration, and the $S_{ndcg}(L)$ measures the relevance of the recommendation list. To put it together, commonly, the state-of-the-art implements the trade-off proposed in \cite{Steck:2018:Calib}. The author's equation balances both as described below:

\begin{equation}\label{eq:tradeoff}
L^* = \underset{ L,\ |L| = N}{\arg\max} (1-\lambda(u)) \cdot S(L) - \lambda(u) \cdot C(P,Q(L)).
\end{equation}

\noindent $L^*$ is the final recommendation list, which is the most optimal. Recent studies have proposed some new formulations. Seymen et al.~\cite{Seymen:2021:Constraint} discuss a new trade-off called Calib-Opt, which is a constraint optimization. da Silva and Durao~\cite{Silva:2022:Protocol} propose to use the user's bias in a logarithmic equation, considering it a list value regulator. Naghiaei et al.~\cite{Naghiaei:2022:Towards} approach a new trade-off balance called $CCL$, which considers the item diversity besides the relevance and calibration.

The $\lambda(u)$, in the trade-off equation, is a weight that varies between $[0, 1] \in \mathbb{R}$. When its value is 0, the relevance function constructs the list. When it is 1, the calibration constructs it. Most studies use constant values for all users, such as $0.0, 0.1, 0.2, 0.3, ..., 1.0$~\cite{Kaya:2019:Intent}. da Silva et al.\cite{Silva:2021:Exploiting} present two ways to find the personalized weight, observing the user's preference to determine its value.

To select the items composing the recommendation list, Steck~\cite{Steck:2018:Calib} indicates implementing a greedy algorithm called Surrogate Submodular. This algorithm works position by position, including the best item in its position. However, other studies propose to use new algorithms. The work \cite{Starychfojtu:2021:Recepies} argues that it is possible to select the items using the Fuzzy logic algorithm called Fuzzy D'Hondt. The studies \cite{Zhao:2020:Distortion, Zhao:2021:Rabbit} implement the Top-Z to select the items. Seymen et al.~\cite{Seymen:2021:Constraint} use the Gurobi branch\&Bound. Abdollahpouri et al.~\cite{Abdollahpouri:2023:MinCost} changed Steck's formulation, proposing to choose the items with Minimum-Cost Flow.

\section{Distribution Derivation}
\label{sec:back}
\begin{figure*}[!ht]
	\centering
	\includegraphics[width=\linewidth]{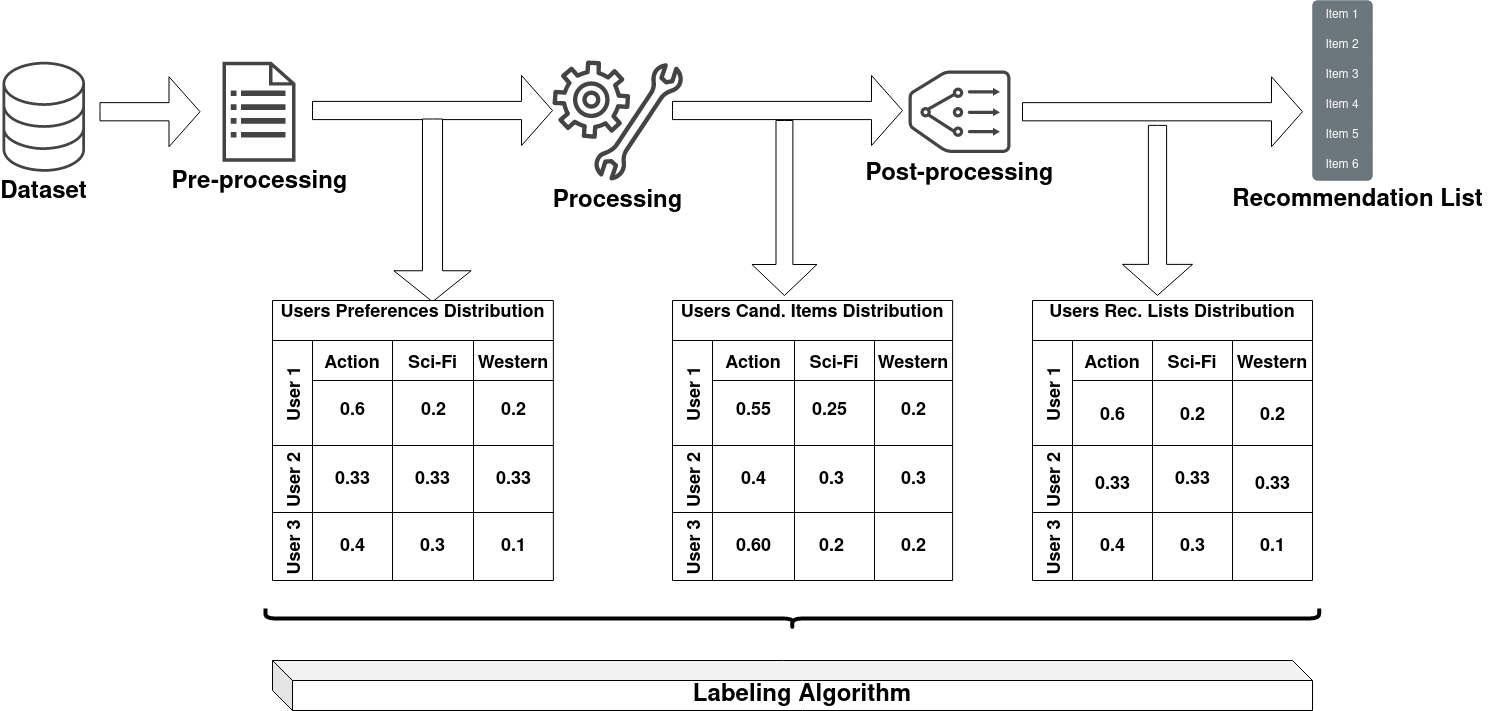}
	\caption{Basic explanation of the proposal.}
	\label{fig:proposal}
\end{figure*}

This section explains the extraction/derivation of the distribution from the data. The example presents a context of movie recommendation using the genre approach and the weighted strategy, similar to those used in \cite{Steck:2018:Calib, Kaya:2019:Intent, Silva:2021:Exploiting}. Thus, Tables \ref{tab:genres_approach_example}, \ref{tab:target_weighted_strategy_example}, and \ref{tab:weighted_distribution} present an example of genre weighted distribution. 

Table \ref{tab:genres_approach_example} shows the computation of the genres' proportion values. The table has three items $i$. Each genre ($g$) in an item ($i$) has its proportion value computed, as shown in the column $p(g|i)$. The column $w_{u, i}$ represents the feedback given by the hypothetical user $u$ over each item $i$. 

\begin{table}[ht!]
	\centering
	\caption{An example of a genre proportion extraction.}
	\label{tab:genres_approach_example}
	\begin{tabular}{rcc}
		\hline
		\textbf{Item}				& \textbf{p(g$|$i)}			& \textbf{$w_{u,i}$}                                                 \\
		\hline
		
		Three Musketeers                  & \begin{tabular}[c]{@{}c@{}}$Adventure=1;p(g|i)=\frac{1}{2}=0.5$\\ $Comedy=1;p(g|i)=\frac{1}{2}=0.5$\end{tabular}          & 5                                                         \\ 
		
		& & \\ 
		The Whale                  & $Drama=1;p(g|i)=\frac{1}{1}=1$                             & 4                                                                                                              \\ 
		
		& & \\ 
		Batman                  & \begin{tabular}[c]{@{}c@{}}$Action=1;p(g|i)=\frac{1}{4}=0.25$ \\ $Adventure=1;p(g|i)=\frac{1}{4}=0.25$\\ $Crime=1;p(g|i)=\frac{1}{4}=0.25$\\ $Drama=1;p(g|i)=\frac{1}{4}=0.25$\end{tabular} & 4 \\ 
		
		\hline
	\end{tabular}\\
\end{table}

Table \ref{tab:target_weighted_strategy_example} has five lines, where each line is a genre that composes the three items represented in the columns. The movie ``Three Musketeers'' scores 5 with two genres, each with a $p(g|i)=0.5$. Applying Equation \ref{eq:weighted_distribution_p}, the numerator for each genre is $w_{u,i}\cdot p(g|i)=5\cdot0.5$. Each movie with the Adventure genre computes the numerator value and is summed in the end, obtaining $3.5$. The denominator is the sum of the scores $5 + 4 = 9$, $5$ from ``Three Musketeers'' and $4$ from ``Batman''. $p(g|i)$ is zero where the genre does not appear. The final value for each class in this example is in the column $p(g|u)$. 

\begin{table}[h!]
	\centering
	\caption{Computation of each genre in each movie.}
	\label{tab:target_weighted_strategy_example}
	\begin{tabular}{ccccc}
		\hline
		\multicolumn{1}{r}{\textbf{Genre}}   & \multicolumn{1}{c}{\textbf{\begin{tabular}[c]{@{}c@{}}Three \\ Musketeers\end{tabular}}} & \multicolumn{1}{c}{\textbf{\begin{tabular}[c]{@{}c@{}}The \\ Whale\end{tabular}}} & \multicolumn{1}{c}{\textbf{\begin{tabular}[c]{@{}c@{}}Batman\end{tabular}}} & \multicolumn{1}{l}{\textbf{p(g$|$u)}} \\ 
		\hline
		
		\multicolumn{1}{r}{$Action$}   & \multicolumn{1}{c}{5$\cdot$0}                                                                       & \multicolumn{1}{c}{4$\cdot$0}           & \multicolumn{1}{c}{4$\cdot$0.25}                                                                            & \multicolumn{1}{l}{1/4=0.25}        \\ 
		
		\multicolumn{1}{r}{$Adventure$} & \multicolumn{1}{c}{5$\cdot$0.5}                                                                     & \multicolumn{1}{c}{4$\cdot$0}           & \multicolumn{1}{c}{4$\cdot$0.25}                                                                            & \multicolumn{1}{l}{3.5/9=0.388}      \\ 
		
		\multicolumn{1}{r}{$Comedy$}    & \multicolumn{1}{c}{5$\cdot$0.5}                                                                     & \multicolumn{1}{c}{4$\cdot$0}           & \multicolumn{1}{c}{4$\cdot$0}                                                                               & \multicolumn{1}{l}{2.5/5=0.5}      \\ 
		
		\multicolumn{1}{r}{$Crime$}     & \multicolumn{1}{c}{5$\cdot$0}                                                                       & \multicolumn{1}{c}{4$\cdot$0}           & \multicolumn{1}{c}{4$\cdot$0.25}                                                                            & \multicolumn{1}{l}{1/4=0.25}        \\ 
		
		\multicolumn{1}{r}{$Drama$}     & \multicolumn{1}{c}{5$\cdot$0}                                                                       & \multicolumn{1}{c}{4$\cdot$1}           & \multicolumn{1}{c}{4$\cdot$0.25}                                                                            & \multicolumn{1}{l}{5/8=0.625}        \\ 
		
		\hline
	\end{tabular}\\
\end{table}

Table \ref{tab:weighted_distribution} shows the distribution values to the $P$, $Q$ and $\tilde{Q}$ distributions. The table has seven lines representing a genre in the class set ($\forall g \in G$). In Lines 6 and 7, the distribution values are zero. It happens because these do not belong to any item in the hypothetical user preference set. 

\begin{table}[h!]
	\centering
	\caption{Final distribution values.}
	\label{tab:weighted_distribution}
	\begin{tabular}{cccc}
		\hline
		\multicolumn{1}{r}{\textbf{Genres}} & \multicolumn{1}{c}{\textbf{$P$}} & \multicolumn{1}{c}{\textbf{$Q$}} & \multicolumn{1}{l}{$\tilde{Q}$} \\ 
		\hline
		\multicolumn{1}{r}{\textbf{Action}}   & \multicolumn{1}{c}{0.25}      & \multicolumn{1}{c}{0.0}        & \multicolumn{1}{l}{0.0025}                 \\ 		
		\multicolumn{1}{r}{\textbf{Adventure}} & \multicolumn{1}{c}{0.388}      & \multicolumn{1}{c}{0.35}        & \multicolumn{1}{l}{0.35038}                 \\ 
		\multicolumn{1}{r}{\textbf{Comedy}}    & \multicolumn{1}{c}{0.5}      & \multicolumn{1}{c}{0.563}        & \multicolumn{1}{l}{0.56237}                 \\ 
		\multicolumn{1}{r}{\textbf{Crime}}     & \multicolumn{1}{c}{0.25}      & \multicolumn{1}{c}{0.4}        & \multicolumn{1}{l}{0.3935}                 \\ 
		\multicolumn{1}{r}{\textbf{Drama}}     & \multicolumn{1}{c}{0.625}      & \multicolumn{1}{c}{0.5}        & \multicolumn{1}{l}{0.50125}                 \\ 
		\multicolumn{1}{r}{\textbf{Romance}}   & \multicolumn{1}{c}{0.0}        & \multicolumn{1}{c}{0.0}        & \multicolumn{1}{l}{0.0}                       \\ 
		\multicolumn{1}{r}{\textbf{Sci-fi}}    & \multicolumn{1}{c}{0.0}        & \multicolumn{1}{c}{0.0}        & \multicolumn{1}{l}{0.0}                       \\ 
		
		\hline
	\end{tabular}\\
\end{table}

Based on the explained distribution derivation, Figure \ref{fig:proposal} shows when and where the distribution can be produced. The first distribution is the user's preference distribution, which is obtained after cleaning the data. The second one is the user's candidate distribution, which is received after the recommender algorithm produces its candidate items, which are usually recommended to the user by traditional recommender systems. The last is the recommendation list distribution obtained after the calibrated method produces its recommendation list. Each user is a line in the distribution, and each column is a genre.

\section{Understanding Distribution Structure}
\label{sec:proposal}
This study aims to understand the distributions used and generated by calibrated recommendation systems from a structural point of view. Our proposal identifies and understands which structure these distributions adhere to and evaluates how similar each user is to another, which makes it possible to observe how many clusters or groups are formed.

Many algorithms can identify the data structure and provide an understanding of it. This study seeks to find the structure using the most varied algorithm types, such as clustering, Gaussian mixture models, neighbor models, ensembles, and outline detectors. In the following, we describe all proposed algorithms for this study.

\begin{itemize}
    \item{ \textbf{K-Means}: It is a well-known algorithm to cluster data. The K-Means \cite{Arthur:2007:SODA} needs as entry the number of clusters - K; the users will be divided, and using this number seeks to produce K clusters well formed. The algorithm considers that some users are centers and are used as a centroid to form the cluster.}
    \item{ \textbf{Fuzzy C-Means (FCM)}: It is an algorithm based on the K-Means. Although the Fuzzy C-Means \cite{Dias:2019:Fuzzy} uses fuzzy logic to find the users' clusters. Using fuzzy logic, a user can be in more than one cluster. However, the algorithm indicates the cluster that the user has more affinity with.}
    \item{ \textbf{Bisecting K-Means (BKM)}: Bisecting K-Means \cite{Di:2018:Bisecting} is another variation of the K-Means. Unlike the original version, it works in the hierarchical division, forming interactive clusters and picking the centroids during the cluster formation.}
    \item{ \textbf{Agglomerative}: It is a hierarchical algorithm that creates the centroids based on a tree structure, known as a dendrogram. In particular, the Agglomerative \cite{Schubert:2021:LWDA} algorithm starts creating one cluster by the user, and each interaction clusters similar users in a cluster until it reaches the number of clusters defined by the entry.}
    \item{ \textbf{Spectral}: It is specialized in finding structure in a sparse matrix. This algorithm works well with a small cluster number and problems based on graphs~\cite{Damle:2018:Spectral}. The Spectral algorithm will help us to understand if the users are divided into a small number of high-density clusters.}
    \item{ \textbf{Birch}: It is an algorithm based on a tree, where subclusters are formed in the leaf, and with the interactions, well-formed clusters are provided \cite{Zhang:1996:BIRCH}.}
    \item{ \textbf{DBScan}: Its algorithm has similarities to the Optics. Both work similarly to find the number of clusters in the data. This algorithm observes the areas with a high density of users and verifies how many dense areas are in the data to set the number of clusters \cite{Schubert:2017:DBSCAN}.}
    \item{ \textbf{Optics}: It is an algorithm that searches for the number of clusters in the data \cite{Ankerst:1999:Optics}. The Optics considers the users as a graph and verifies if it can reach neighboring users based on the distance between the distribution.}

    \item{ \textbf{Gaussian Mixture (GM)}: Its algorithm uses the Bayesian Information Criterion to understand the data and discover the number of clusters. The covariance is used to draw different options of ellipsoids, which define the users' groups \cite{YANG20123950}.}
    \item{ \textbf{Bayesian Gaussian Mixture (BGM)}: It is similar to the Gaussian Mixture. However, it uses variational inference \cite{Roberts:1998:Bayesian}.}

    \item{ \textbf{Isolation Forest (IF)}: As our context, its algorithm performs efficiently in high-dimensional datasets. The Isolation Forest (IF) is a variation of the Random Forest model, observing the data randomly and isolating \cite{Liu:2008:IF}.}
    \item{ \textbf{One-Class SVM (OCSVM)}: Its algorithm is a variation of the Support Vector Machines used to estimate the support in high-dimensional distributions. This algorithm uses the probability of finding a regular structure to set observations of outside users \cite{scholkopf2001estimating}, seeking to form groups based on outliers.}
    \item{ \textbf{One-Class SVM using Stochastic Gradient Descent (OCSVMSGD)}: It is a variation of One-Class SVM that works in a linear version and can scale linearly.}
    \item{ \textbf{Elliptic Envelope (EE)}: Its algorithm creates ellipses based on central data users, considering that the users that are not in the ellipses are outliers, i.e., from other groups \cite{Peter:1999:Envelope}.}
    \item{ \textbf{Local Outlier Factor (LOF)}: It is an outlier detector based on k-nearest neighbors. LOF uses the density deviation to measure how isolated the user is from the neighbors \cite{Breunig:2000:LOF}.}
\end{itemize}

\section{Methodology}
Following prior work~\cite{Kaya:2019:Intent, Silva:2021:Exploiting, Souza:2024}, we implemented Singular Value Decomposition (SVD) as our base recommendation algorithm (i. e. traditional recommender), chosen for its wide adoption and proven efficiency in collaborative filtering tasks.

The first stage of our recommendation process involved generating a candidate set for each user. This set consisted of 100 items, selected by identifying the 100 items with the highest SVD-predicted ratings from all items the user had not previously interacted with. This candidate set then served as the input for the subsequent calibration step, from which the final top-10 recommendation list was produced. For this calibration, we use 11 different constant values of $\lambda \in [0.0; 0.1; ...; 0.9; 1.0]$, which are represented in our results as $C@0.0, C@0.1 ... C@1.0$.

We employed statistical sampling for the selection of hyperparameters in our SVD recommender algorithm, focusing on four key parameters. For both \textit{n\_factors} (the number of latent factors) and \textit{n\_epochs} (the number of iterations of the SVD procedure), we utilized a uniform integer distribution, sampling random values between 10 and 150. As for \textit{lr\_all} (the learning rate for all parameters) and \textit{reg\_all} (the regularization term for all parameters), we applied a continuous uniform distribution. This allowed the learning rate to vary between 0.001 and 0.01, and the regularization term between 0.01 and 0.1. For evaluation, we use a 5-fold cross-validation methodology.

Additionally, we performed hyperparameter tuning on the algorithms used to identify the intrinsic structure of the data. For this tuning, a grid search methodology with 5-fold cross-validation was used. The following list details each hyperparameter, presenting its function, the explored search space, and the algorithms to which it applies.
\begin{itemize}
    \item \textbf{n\_clusters}: This parameter defines the desired number of clusters (or partitions) that the algorithm should identify in the dataset. \textbf{Search space:} 2, 3, 5, 7, 11, 13, 17, 19, 23, 29, 31, 37, 41, 43, 47, 53, 59, 61, 67, 71, 73, 79, 83, 89, and 97. \textbf{Applicable in:} K-Means, Biclustering K-Means, Agglomerative, Spectral, Fuzzy C-Means, and Birch.

    \item \textbf{n\_components}: For probabilistic mixture models such as Gaussian Mixture Models, this parameter specifies the number of individual Gaussian (or other distribution) components to be fitted to the data. Each component represents a sub-population or a latent cluster within the dataset. \textbf{Search space:} 1, 2, 3, 5, 7, 11, 13, 17, 19, 23, 29, 31, 37, 41, 43, 47, 53, 59, 61, 67, 71, 73, 79, 83, 89, and 97. \textbf{Applicable in:} Gaussian Mixture and Bayesian Gaussian Mixture.

    \item \textbf{epsilon}: This parameter specifies the maximum distance between two data points for them to be considered part of the same neighborhood. \textbf{Search space:} 0.05, 0.10, 0.15, 0.20, 0.25, 0.3, 0.35, 0.4, 0.45, 0.,5 and 0.55. \textbf{Applicable in:} DBScan and Optics.

    \item \textbf{min\_samples}: Essential for density-based algorithms like DBSCAN and OPTICS. It defines the minimum number of data points (including the point itself) required within a neighborhood (defined by \texttt{epsilon}) for a point to be considered a ``core point" of a cluster. \textbf{Search space:} 2, 3, 5, 7, 11, 13, 17, 19, 23, 29, 31, 37, 41, 43, 47, 53, 59, 61, 67, 71, 73, 79, 83, 89, and 97. \textbf{Applicable in:} DBScan and Optics.

    \item \textbf{metric}: This parameter determines the distance or similarity measure used by the algorithm to calculate the proximity between data points. These metrics influence how the algorithm perceives the ``closeness'' of points. \textbf{Search space:} cityblock, cosine, euclidean, L1, L2, manhattan, braycurtis, canberra, chebyshev, correlation, and hamming. \textbf{Applicable in:} DBScan, Optics and Local Outlier Factor.

    \item \textbf{n\_estimators}: Predominantly used in ensemble methods. It represents the number of individual estimators that are built and combined to form the final model. More estimators generally lead to a more robust model but increase computational cost. \textbf{Search space:} 1, 2, 3, 5, 7, 11, 13, 17, 19, 23, 29, 31, 37, 41, 43, 47, 53, 59, 61, 67, 71, 73, 79, 83, 89 and 97. \textbf{Applicable in:} Isolation Forest.

    \item \textbf{n\_neighbors}: Specifically used by neighborhood-based algorithms. This parameter specifies the number of nearest neighbors to consider when calculating the local density or outlier score of a data point. It directly impacts the local context considered for anomaly detection. \textbf{Search space:} 1, 2, 3, 5, 7, 11, 13, 17, 19, 23, 29, 31, 37, 41, 43, 47, 53, 59, 61, 67, 71, 73, 79, 83, 89 and 97. \textbf{Applicable in:} Local Outlier Factor.

    \item \textbf{nu}: A crucial parameter for One-Class SVM (OCSVM) and related anomaly detection algorithms. It represents an upper bound on the fraction of training errors (outliers) and a lower bound on the fraction of support vectors. Essentially, it controls the trade-off between the number of detected outliers and the tightness of the decision boundary. \textbf{Search space:} 0.05, 0.1, 0.2, 0.3, 0.4 and 0.5. \textbf{Applicable in:} One-Class SVM, One-Class SVM using Stochastic Gradient Descent and Elliptic Envelope.
\end{itemize}

In our experiments, we processed the user-item scores using two distinct approaches:
\begin{itemize}
    \item \textbf{Original score:} The original user ratings were used directly, preserving their native scale (e.g., 0-5 or 0-10).
    \item \textbf{Binary score:} The original ratings were transformed into binary values based on the dataset's scale. For scales from 0 to 5, a score of $\geq 4$ was mapped to 1 (and 0 otherwise). For scales from 0 to 10, a score of $\geq 8$ was mapped to 1 (and 0 otherwise).
\end{itemize}

\section{Experimental Setup}
\label{sec:setup}
We conducted offline experiments to demonstrate the effectiveness of our proposal for identifying the users' distribution structures.

\subsection{Datasets}
We adopted three datasets from the movie domain. The first dataset is the Movielens 1M~\cite{Harper:2015:Movielens}. The second one is Yahoo Movies~\footnote{https://webscope.sandbox.yahoo.com/catalog.php?datatype=r\&did=4}. The third one is Twitter Movies~\footnote{https://www.kaggle.com/datasets/tunguz/movietweetings}. The dataset numbers before and after applying the pre-processing are presented in Table \ref{tab:datasets}. We clean and filter the data using the following rules: i) We remove all items without genre information; ii) We drop all users that have less than 50 transactions, due to our 5-cross-validation methodology, and 10 items in the recommendation list; and iii) We eliminate all items that have no interaction with any user. 

\begin{table}[th!]
	\centering
	\caption{Datasets before and after the pre-processing step.}
	\label{tab:datasets}
	\begin{tabular}{lcccc}
		\hline
		\textbf{Datasets}                 & $|U|$     & $|I|$  & $|W|$   & $|G|$ \\ \hline
		Original ML1M                     & 6040      & 3883   & 1000209 & 18  \\
		\textbf{Cleaned ML1M}             & 4247      & 3883   & 940971  & 18  \\
		Original Yahoo Movies             & 7642      & 62423  & 221367  & 20  \\
		\textbf{Cleaned Yahoo Movies}     & 270       & 2439   & 30577   & 19  \\ 
		Original Twitter Movies           & 70783     & 37342  & 906831  & 29  \\
		\textbf{Cleaned Twitter Movies}   & 3836      & 37268  & 529668  & 28  \\\hline
	\end{tabular}
\end{table}

Each cleaned dataset was divided into five fixed folds to allow for the 5-cross-validation methodology. Thus, the following results are an average of five runs.

\subsection{Metrics}
This study proposes to understand the distributions in a calibrated recommendation system. Yet, there are no previous definitions of better structures or clusters. Thus, to evaluate our proposal, we implement metrics that evaluate the system results without known cluster labels. All metrics are described below:

\begin{itemize}
    \item{ \textbf{Silhouette}: A well-known metric \cite{ROUSSEEUW:1987:JCAM} used when the clusters and their members are not truly known. Thus, the data defines the similarity and proximity along the points inside and outside the clusters. The a(i) is the average dissimilarity of $i$ to all other objects in its cluster. The b(i) is the average dissimilarity of $i$ to all other objects outside its cluster. The silhouette range is between -1 and 1.}
    \begin{equation}\label{eq:silhouette}
        sil(i) = \frac{b(i) - a(i)}{max\{a(i), b(i)\}}.
    \end{equation}
    
    \item{ \textbf{Jaccard Score}: This metric scores the number of items in A and B. A and B are arrays where the index is the item, and the value is the cluster identification. A is the clusters from the users' preferences distribution. B is the clusters from the users' candidate items distribution or recommendation lists distribution. Thus, this metric helps us understand whether users change the cluster during the recommender algorithm or the calibration step. The users are expected to belong to the same cluster after all stages.}
    \begin{equation}\label{eq:jaccard}
    J(A,B) = \frac{|A \cap B|}{|A \cup B|}
    \end{equation}
\end{itemize}

Besides these two metrics, we use the Mean Average Precision (MAP), Mean Reciprocal Rank (MRR), and Mean Average Calibration Error (MACE), similar to \cite{Silva:2021:Exploiting, Silva:2022:benchmark}.

\subsection{Experimental Results}
\label{sec:results}
We now discuss the experimental results from the two experiments (binary and original) in the three datasets and the fifteen labeling algorithms evaluated by five metrics. We here first provide a short overview of these results without differentiating between research questions, and then we explain the results in detail, given the six research questions.

\begin{figure*}[!ht]
    \centering
    \includegraphics[width=.8\linewidth]{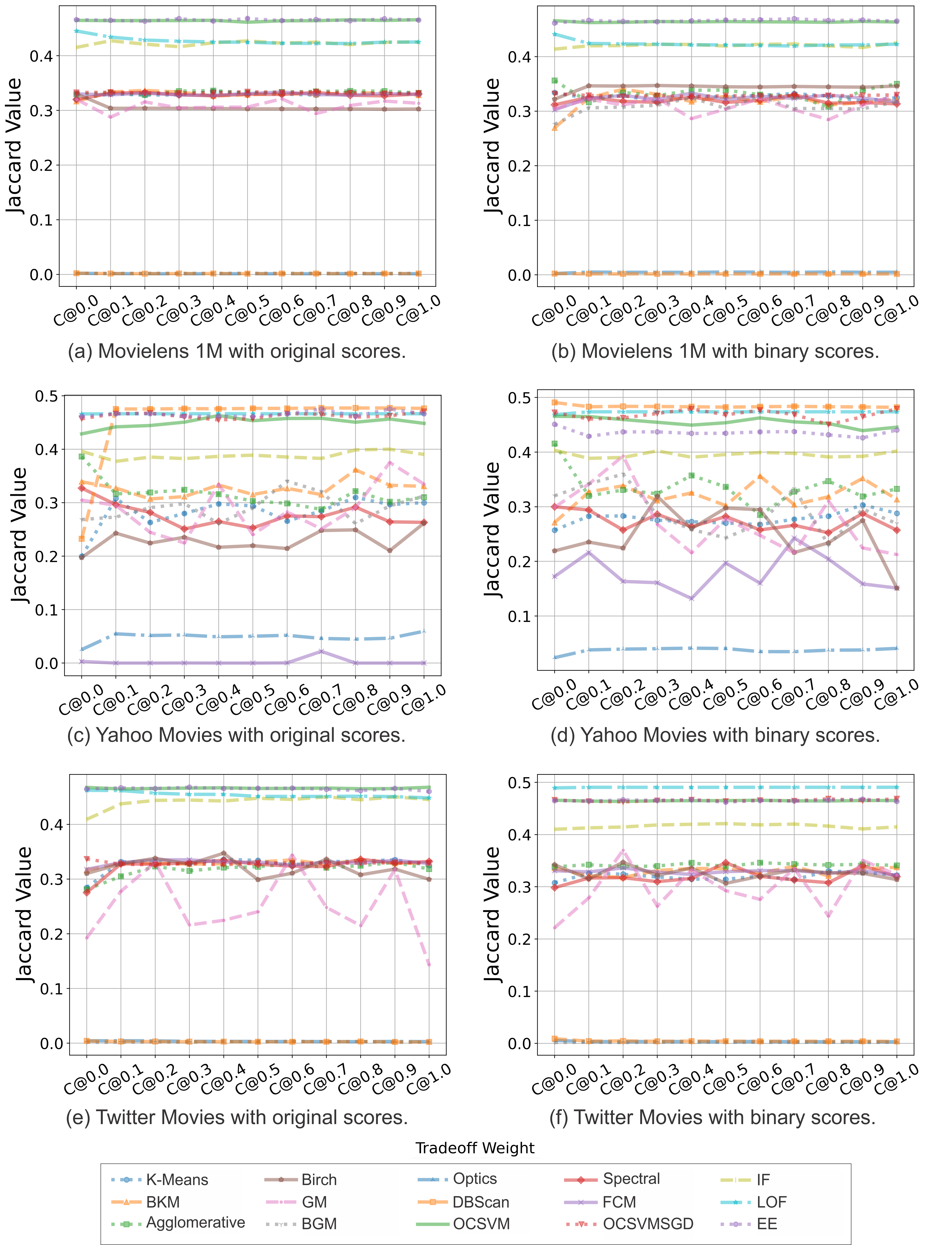}
    \caption{This figure presents the Jaccard metric, illustrating the degree of user group changes. The first x-point $C@0.0$ represents the distribution of candidate items, and the subsequent points $C@0.1$ to $C@1.0$ show the calibrated recommendation list distribution as the tradeoff weight varies.}

    \label{fig:jaccard}
\end{figure*}

\begin{figure*}[!ht]
    \centering
    \includegraphics[width=.8\linewidth]{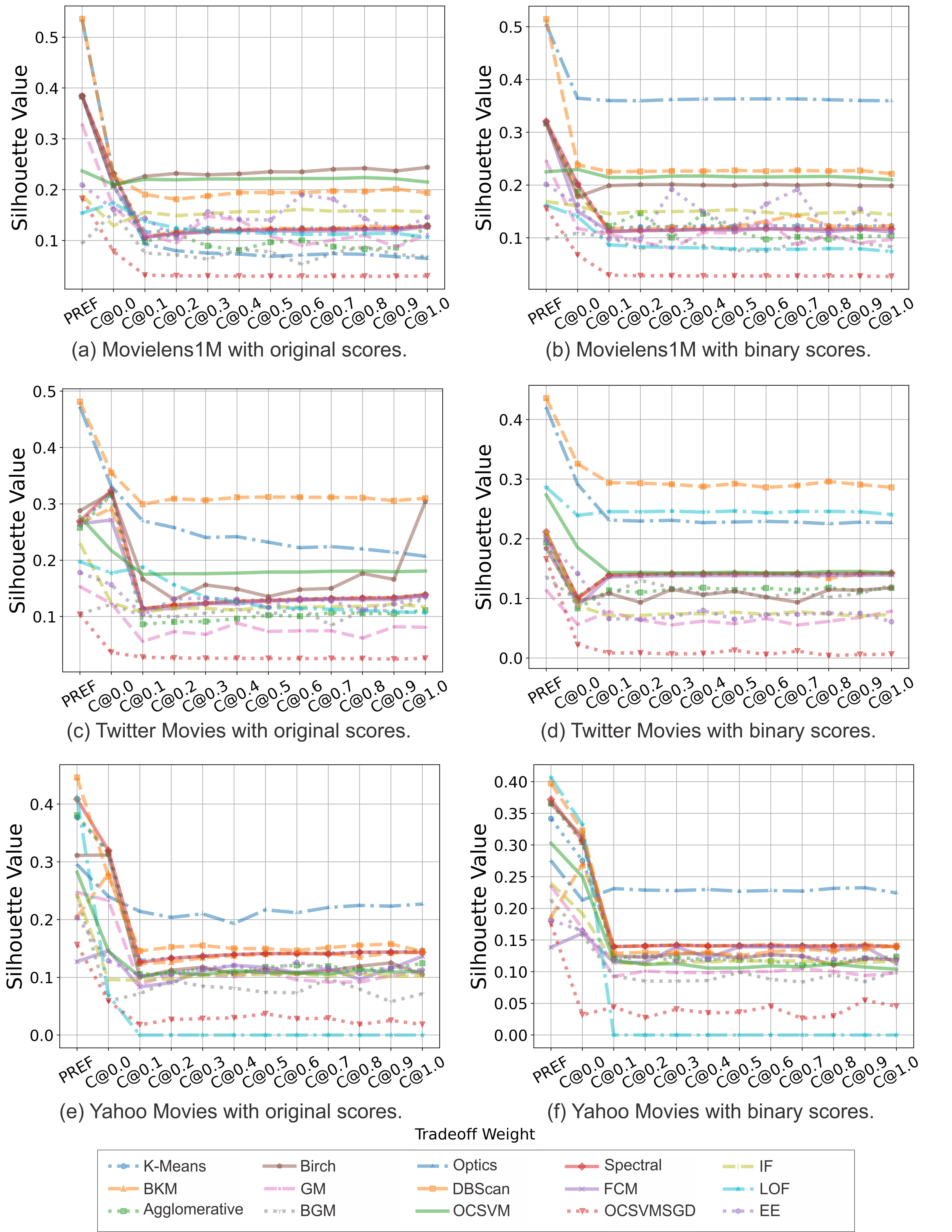}
    \caption{This figure presents the silhouette measure, indicating how well the distributions adhere to identified structures. The first x-point (PREF) corresponds to the users' preferences distribution, the second x-point $C@0.0$ represents the candidate items distribution, and subsequent points $C@0.1$ to $C@1.0$ show the calibrated recommendation list distribution as the tradeoff weight varies.}

    \label{fig:silhouette}
\end{figure*}

Figure \ref{fig:jaccard} shows the Jaccard similarity between the top-100 items learned at $C@0.0$ (the uncalibrated SVD ranking) and each calibrated list as $\lambda$ value increases. A high Jaccard value means the calibrated ranking still contains most of the original items. A low value means calibration is severely reordering or replacing films. Figures \ref{fig:jaccard} and \ref{fig:silhouette} reveal the optimal trade-off where we maintain cluster structure and list stability, where the small-to-moderate values of $\lambda$ introduce calibration without wholesale disruption.

The mean silhouette coefficient presented in Figure \ref{fig:silhouette} shows that for each of our fifteen structure‐learning methods, as the calibration weight $\lambda$ varies from 0 (pure SVD) to 1 (full historical‐profile matching). This curve directly addresses RQ 1—how calibration reshapes the latent clusters among users’ genres. High silhouette values (closer to +1) indicate that a method still finds tight, well‐separated groups even after blending in the user’s true preference distribution. By comparing these curves side by side, we can see which algorithms (for example, DBSCAN or Gaussian Mixtures) remain robust to moderate calibration and which quickly lose their internal cohesion as $\lambda$ grows.

Despite using fifteen different structure‐learning methods, the absolute silhouette values in Figure \ref{fig:silhouette} remain in the weak to moderate range (mean $s\approx0.2$–$0.4$). This is not surprising, for two key reasons. First, our genre-vector space is high-dimensional and smooth. Second, as the calibration weight $\lambda$ increases, each user’s calibrated distribution $\widetilde Q$ becomes a more uniform blend of their historical tastes, further blurring natural cluster boundaries. In such settings, even a well-separated algorithm like Gaussian Mixtures will yield modest silhouette scores, because the inter-cluster distances shrink relative to within-cluster variance.

Importantly, a low silhouette here does not imply that calibration has failed, only that our discrete clustering abstractions struggle to carve crisp segments from a continuum of individualized preferences.  We therefore interpret silhouette trends, the shape of each method’s curve, as more informative than the raw magnitudes.  A method that maintains or even increases its silhouette for small $\lambda$ (e.g.\ DBSCAN up to $\lambda\approx0.3$) is one whose cluster assignments are robust to modest calibration. Conversely, methods whose silhouette decreases as soon as $\lambda>0$ are overly sensitive to distributional smoothing and may not be suitable for applications requiring both fidelity and interpretability of user‐taste groupings.

\begin{figure*}[!ht]
    \centering
    \includegraphics[width=.8\linewidth]{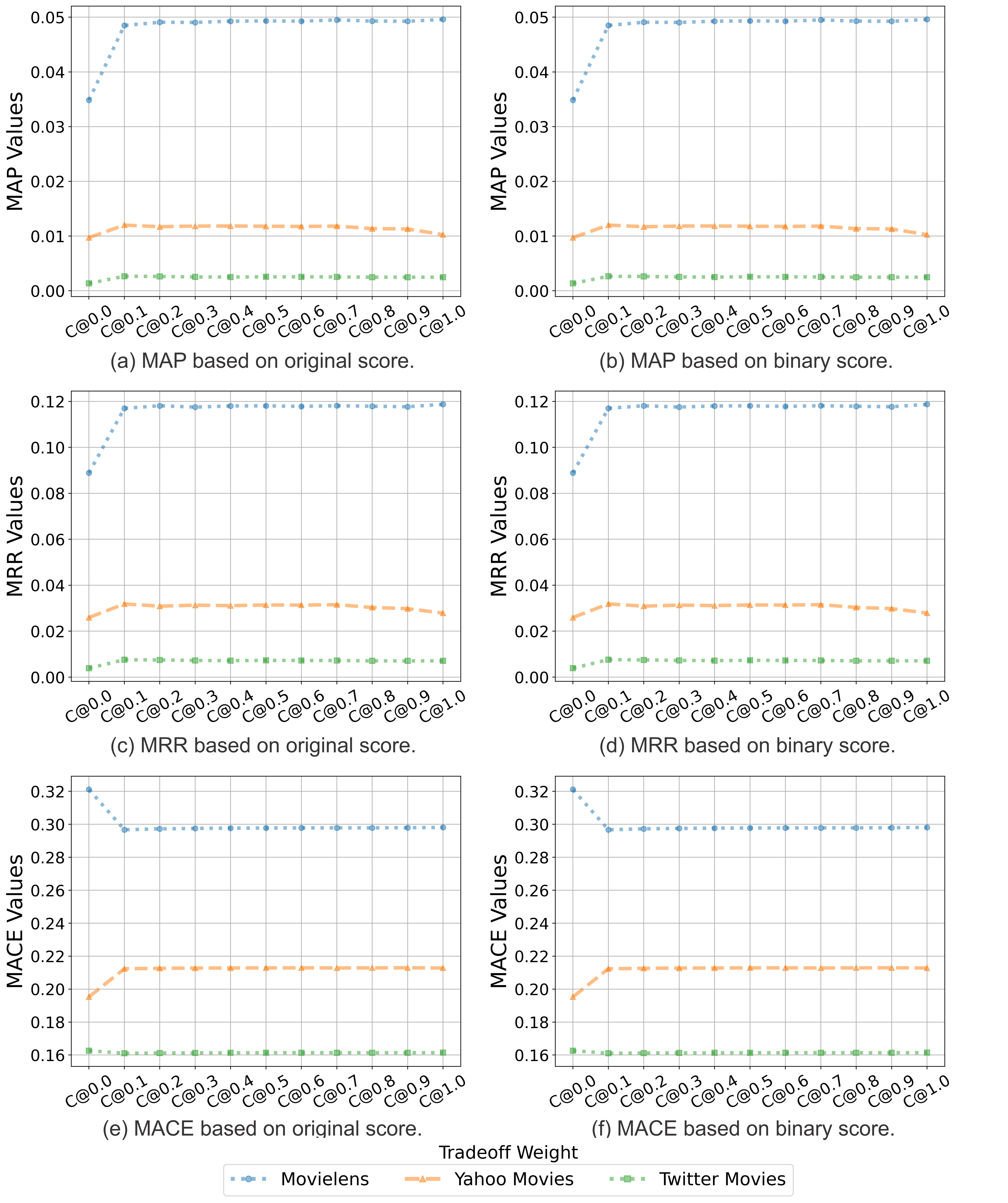}
   \caption{This figure presents the recommender system metrics MAP, MRR, and MACE. The first point on the x-axis ($C@0.0$) represents the distribution of the candidate items, while the subsequent points correspond to the distributions of the calibrated recommendation lists under varying trade-off weight values. The figures in the first row show the results using the original scores, whereas the figures in the second row use binary scores.}

    \label{fig:all:datasets}
\end{figure*}

\subsection{\textbf{RQ1}) How many groups are present in the user distribution patterns?}

Most models found two clusters or groups for all datasets in both experiments (binary and original scores). To achieve this value, all algorithms are optimized through their hyperparameter, seeking the best silhouette. Each model is proposed to fit the three distributions, and the silhouette metric is computed for each distribution. 

Most cluster algorithms need as an entry the number of clusters in which the data will be divided. Thus, we ran a search using the $24$ prime numbers between $2$ and $89$ for these algorithms. Unlike other clustering algorithms, OPTICS and DBSCAN do not require the number of clusters to be specified as a hyperparameter, as they are designed to discover the number of clusters automatically based on the data distribution. Thus, we ran a combination of $1452$ hyperparameters on these two algorithms. However, both models found fewer clusters than the other clustering algorithms. They also found different clusters for user preferences, candidate items, and recommendation list distributions. These findings happen in both experiments. For instance, in Movielens, the Optics found in the users' preferences $102$ clusters, the candidate items $217$ clusters, and the recommendation list $182$ clusters. Its effect can be observed in Figure \ref{fig:jaccard}, where for Movielens, Yahoo Movies, and Twitter Movies, the Optics and DBScan achieve $0$ in the Jaccard metric because, in the two experiments, the number of clusters found in each distribution is different. In Figure \ref{fig:jaccard} Yahoo Movies (c-d), the Bayesian model also found different clusters between the distributions.
Like most clustering algorithms, all other implemented models divide the users into two groups or labels. This result indicates that in the movie domain, the users are divided based on the number of genres they prefer. Thus, unlike the division made by \cite{Abdollahpouri:2019:Unfairness, Naghiaei:2022:UnfairnessBook}, which considers three popularity groups, the G-dimensionality based on the genres divides the users into those focused on fewer genres and adept at many genres.

\subsection{\textbf{RQ2}) Do users switch groups when comparing the inherent distributions of candidate items and calibrated recommendations with their original preference distributions?}
Yes. The Jaccard metric comprises the degree of change in the data in one value. It indicates if the user's candidate items or calibrated recommendation list are cohesive with their preferences. Figure \ref{fig:jaccard} shows the results from the three datasets used in the two experiments. More than half of the users have changed groups during the workflow. It happens in all experiments for all datasets. It is possible to observe that the same algorithms achieve better results. For Movielens shown in Figure \ref{fig:jaccard} (a-b), the best results are with the Envelope and OSVM. The LOF and IF follow them, achieving the second-best results. These four algorithms are outlier detectors. For Yahoo Movies, Figure \ref{fig:jaccard} (c-d), the best results are with the DBScan, followed by the LOF and SGD. Other algorithms that achieve high values are the OSVM and Envelope. LOF, SGD, OSVM, and Envelope are outlier detectors. Twitter Movies' best results are with the LOF, OSVM, and SGD. Another algorithm that achieves a high value is the IF. Thus, based on the results, we can infer that the outlier detector models are the best for observing the changes in labels.

The Optics achieve the worst results for all datasets, changing all users from their initial label. For Movielens and Twitter Movies, the DBScan achieves the worst results with the Optics. We discuss it in the \textbf{RQ1}. All other algorithms exhibit similar behavior and achieve intermediate results. The results are divided into three blocks, as shown in Figure \ref{fig:jaccard}, better, intermediate, and worst results.

It is possible to observe that the calibrated recommendations context does not provide recommended items that force the users to change the group more than the recommendation list supplied by the traditional recommender SVD. This finding indicates that the calibrated recommendation list aligns with the user's preferences to the same degree and, in high-dimensional problems, does not produce a miscalibrated distribution. It is essential to note that the calibrated recommendation is tolerable with some degree of miscalibration, as the user adds new preferences with the system's use. Therefore, the system will adapt and begin to incorporate some elements of this new genre. This action allows the user to move among groups. However, as debated by the state-of-the-art \cite{Silva:2023:Novel}, calibration can improve many objectives. For instance, it creates a more personalized recommendation list and increases the number of serendipitous items. Thus, changing the user groups during the recommendation does not result in a worse performance.

\subsection{\textbf{RQ3}) What kind of structure do user distributions follow?}
The silhouette metric indicates which structure the distributions are adherent to. Figure \ref{fig:silhouette} shows that the three datasets adhere to the cluster structure. It is possible to verify that the Optics and DBScan algorithms fit the best silhouettes for most datasets and experiments. In special, the Optics fits better on Yahoo Movies. The DBScan achieves the best silhouette for Twitter Movies, followed by the LOF (neighbor-based) and the Optics. Unlike the Twitter Movies algorithm, the LOF algorithm yields one of the worst results for Movielens and the worst for Yahoo Movies. The worst adherence is with the SGD model. This algorithm achieves the worst result in both Movielens and Twitter Movies experiments. The algorithms that accomplish the intermediate results obtain close silhouette values.

The users' preferences distribution achieves a higher silhouette than the distributions of their candidate items and recommendation lists for all datasets in both experiments. This behavior happens due to the amount of data in the users' profiles. The users' candidate items and recommendation list distributions each contain ten items. Thus, the models can not fit a better silhouette for this amount of data regarding the users' preferences.

The silhouette value indicates that most algorithms identify a poor structure because its value is lower than $0.5$. There is an exception where the silhouette is reasonable, greater than $0.5$, achieved by the algorithm Optics and DBScan in Movielens.

\subsection{\textbf{RQ4}) Do different score types (e.g., binary vs. continuous) lead to different distribution structures?}
For most of the results, no. It is possible to observe in Figure \ref{fig:silhouette} that the algorithms' silhouette values are similar. However, there are some differences. In the Movielens, Figure \ref{fig:silhouette} (b) shows that Optics and DBScan fit a better silhouette than the other algorithms to the calibrated recommender distribution. Although Figure \ref{fig:silhouette} (a) shows that the Birch and OSVM fit it better. Some variations are observable in Figures \ref{fig:silhouette} (e) and \ref{fig:silhouette} (f), where the LOF achieves the second-best silhouette in one experiment and then drops to an intermediary performance in the other. 

\subsection{\textbf{RQ5}) Which type of score results in better metric values?}
There is no difference between the score types. It is possible to observe in Figure \ref{fig:all:datasets} that all the metrics achieve the same values for all datasets. It indicates that using binary or original scores does not affect the results. The experiment considers the score transformation in the pre-processing.

\subsection{\textbf{RQ6}) Can distribution structures be used to evaluate calibrated systems?}
The results demonstrate that it is possible to utilize clustering, grouping, or labeling algorithms to assess the degree of miscalibration in a calibrated recommendation system context. Figure \ref{fig:jaccard} presents the degree of users in the same preference group after receiving a calibrated recommendation. Its degree is the same as that of the traditional recommender algorithm SVD implemented in this study as a candidate items producer and baseline. In Figure \ref{fig:silhouette}, it is possible to observe that ten items in the candidate items set and recommendation list do not provide sufficient items to identify a better silhouette. Across all trade-off weights, the values remain similar or identical. However, at least 50 user preference items can provide a better silhouette identification.

\subsection{Discussion}

Our investigation aimed to unveil the underlying structures of user preference distributions in calibrated recommendation systems, an aspect underexplored in previous works. By employing 15 diverse algorithms, including clustering and outlier detection methods, and by analyzing three key distributions (user preferences, candidate items, and the calibrated recommendation list), we obtained significant insights that answer our six research questions (RQ1-RQ6).

The results consistently demonstrate that user preference distributions in the movie domain tend to cluster into two main patterns, differing from previous unidimensional categorizations based on popularity. These two groups can be understood as ``specialists'' (users focused on a few categories) and ``generalists'' (users with broader preferences). This G-dimensional distinction (based on genres) offers a new perspective on user categorization, overcoming the limitations of unidimensional analyses.

A crucial finding, addressed in RQ2, is that more than half of the users switch behavioral groups when comparing their original preference distributions with the distributions of candidate items or calibrated recommendation lists. Interestingly, this shift is no more pronounced in the context of calibrated recommendations than in lists generated by traditional systems like SVD. This suggests that the calibrated system, by aligning with user preferences, does not force an over-specialization that would lock them into a single group, but rather allows for a tolerance to a certain degree of miscalibration. This tolerance is beneficial, as it enables the system to adapt to new user preferences, which can lead to more personalized lists and an increase in unexpected yet relevant items. Outlier detection models (such as Elliptic Envelope, Isolation Forest, One-Class SVM, and Local Outlier Factor) proved to be the most effective in identifying these label changes, surpassing traditional clustering algorithms in observing these dynamics.

Regarding the structure of the distributions (RQ3), the Silhouette coefficient indicated that the distributions adhere to a cluster structure. Algorithms like OPTICS and DBSCAN frequently achieved the best silhouette scores, suggesting they are more adept at identifying the characteristics of these distributions. However, most algorithms identified a ``poor structure'' (silhouette generally below 0.5), a limitation that warrants attention. This may be attributed to the high dimensionality (G-dimensional) and the fluid, overlapping nature of user taste profiles, which complicates the formation of distinct cluster segments. The greater amount of data in the user preference profiles resulted in better silhouette scores.

A significant conclusion, addressed in RQ4 and RQ5, is that the type of rating used (original vs. binary) does not materially affect the structure of the resulting distributions or the performance of the metrics (Silhouette, Jaccard, MAP, MRR, and MACE). This finding is crucial as it validates the findings of previous studies that employed different rating strategies (e.g., \cite{Steck:2018:Calib, Kaya:2019:Intent} with binary ratings; \cite{Silva:2021:Exploiting, Silva:2022:benchmark} with original ratings), suggesting that their conclusions were not influenced by the choice of rating format.

Finally, the study demonstrates that the analysis of distribution structure can, in fact, be used as a basis for evaluating calibrated recommendation systems (RQ6). Although there are challenges in identifying strong structures in short-list distributions, the ability to identify and track users' adherence to preference groups offers a new and complementary dimension of evaluation beyond traditional accuracy and calibration metrics. The fact that outlier detection algorithms are more effective at capturing distributional shifts points to their relevance in adaptive calibration tasks.

\section{Conclusions and Future works}
\label{sec:final}

In this paper, we have investigated the latent structure of three key distributions in calibrated movie recommendation systems: the historical user-preference distribution, the SVD‐based candidate‐item distribution, and their convex combination under varying calibration weights $\lambda$. By applying fifteen clustering and outlier‐detection methods and evaluating them with both silhouette coefficients and Jaccard similarity to the uncalibrated lists, we demonstrated that users’ genre vectors naturally partition into two broad groups, one representing genre specialists and the other generalists, and that this two‐cluster structure persists under moderate calibration $\lambda\leq 0.3$. As $\lambda$ increases beyond this threshold, cluster cohesion degrades, silhouette drops, and the overlap between recommendations and the original SVD ranking diminishes, Jaccard falls, indicating the limits of blending relevance with fidelity. Notably, outlier-detection algorithms proved more robust than hard-clustering approaches in capturing distributional shifts, suggesting they may be better suited for adaptive calibration tasks.

Looking forward, we plan to extend this framework along several dimensions. First, we will explore alternative distribution definitions, such as popularity‐weighted or temporal subprofile distributions, and richer extraction strategies (e.g.\ weighted probability smoothing) to capture nuanced user tastes. Second, we aim to evaluate our methodology in additional domains beyond movies (e.g.\ music, books, points of interest) and with streaming, session‐based recommenders to assess its generality. Third, we will incorporate advanced structure‐learning techniques, such as spectral clustering, hierarchical methods, dynamic density estimation, and integrate user metadata (demographics, context) to refine calibration. Finally, we intend to conduct user studies to measure the perceptual impact of calibrated recommendations on satisfaction and engagement, closing the loop between offline distribution analysis and real‐world user experience.

\subsection{Limitations}
Despite the obtained results, it is important to highlight some inherent limitations of this study:

\begin{itemize}
    \item \textbf{Number of formed groups:} Our experiments did not achieve good results when using more than two groups in the clustering. This forced us to work with less specific user groups in terms of distribution, which, in turn, limited our ability to extract more detailed and in-depth patterns for each group.

    \item \textbf{Silhouette values below 50\%:} The clustering machine learning algorithms employed were not able to define a silhouette greater than 50\%. This result significantly hindered the adjustment of the models and the identification of a strong structure in the data, which could have provided more robust insights.
    
    \item \textbf{Limitation to the movie domain:} The developed approach was experimentally validated exclusively within the movie domain. Consequently, we cannot claim that the observed patterns or the effectiveness of the methodology can be generalized to other domains, such as music, food, or e-commerce, without additional investigations.
\end{itemize}


\section*{Acknowledgments}
This study was financed by Coordenação de Aperfeiçoamento de Pessoal de Nível Superior - Brasil (CAPES) - Finance Code 88887.685243 / 2022-0

\bibliographystyle{IEEEtran}
\bibliography{IEEEabrv, references}

\newpage

\section{Biography Section}
 
\begin{IEEEbiography}[{\includegraphics[width=1in,height=1.25in,clip,keepaspectratio]{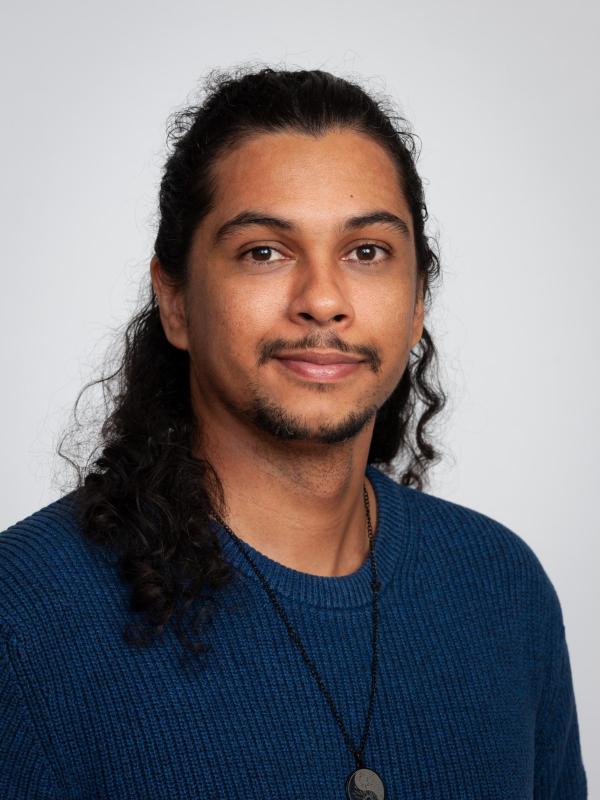}}]{Diego Corr{\^e}a da Silva}
Diego Corr{\^e}a da Silva is a Ph.D. in Computer Science at the Federal University of Bahia. He holds B.Sc. and M.Sc. degrees in Computer Science from the Federal University of Bahia. His main research interests include recommender systems, machine learning, smart homes, fairness, and decision making. Contact him at diego.correa@ufba.br or diego.correa@aau.at.
\end{IEEEbiography}

\begin{IEEEbiography}[{\includegraphics[width=1in,height=1.25in,clip,keepaspectratio]{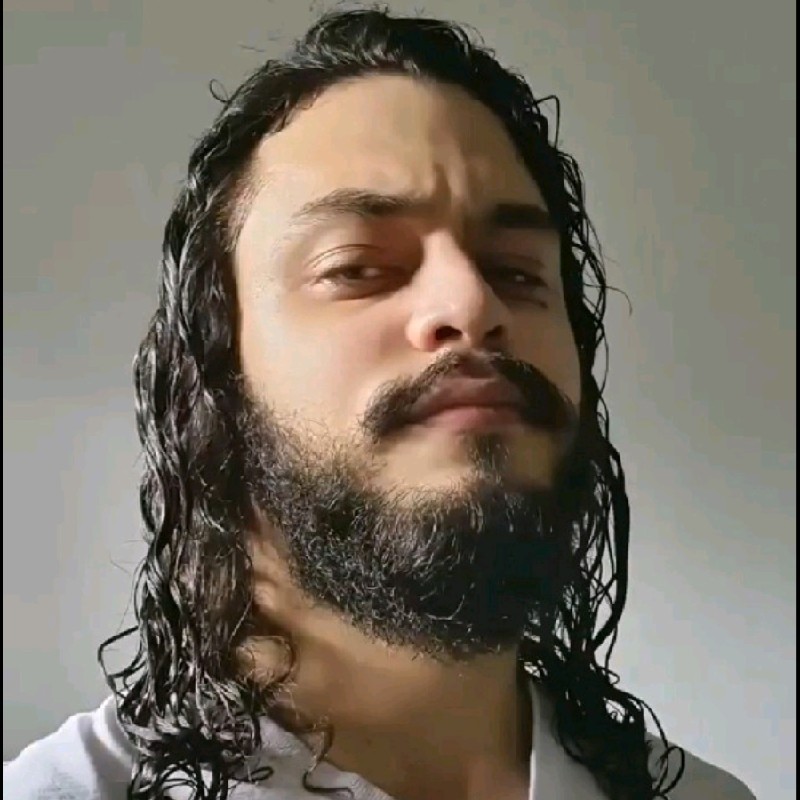}}]{Denis Robson Dantas Boaventura}
Denis Boaventura is a researcher and Ph.D. student in Computer Science at the Federal University of Bahia. He received his B.Sc. in Information Systems from the State University of Bahia and his M.Sc. in Computer Science from the Federal University of Bahia. His research interests include recommender systems, machine learning, smart environments, reinforcement learning, attention, and transformers. Contact him at denis.boaventura@ufba.br.
\end{IEEEbiography}

\begin{IEEEbiography}[{\includegraphics[width=1in,height=1.25in,clip,keepaspectratio]{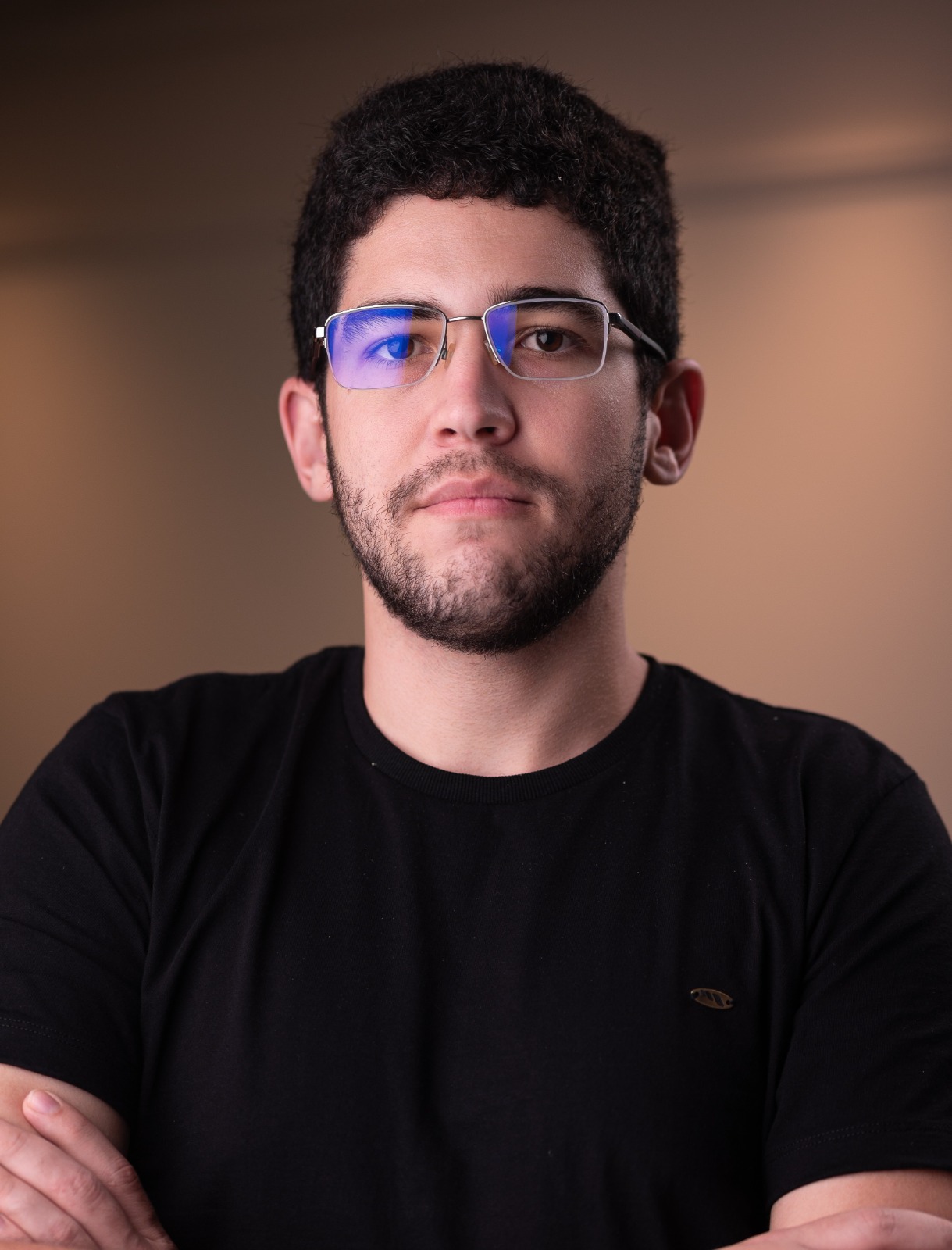}}]{Mayki dos Santos Oliveira}
Mayki dos Santos Oliveira is a PhD student in Computer Science at the Federal University of Bahia (UFBA), where he participates in RECSYS Labs. His main research areas include recommender systems, machine learning, and smart homes. Contact him at maykioliveira@ufba.br.
\end{IEEEbiography}

\begin{IEEEbiography}[{\includegraphics[width=1in,height=1.25in,clip,keepaspectratio]{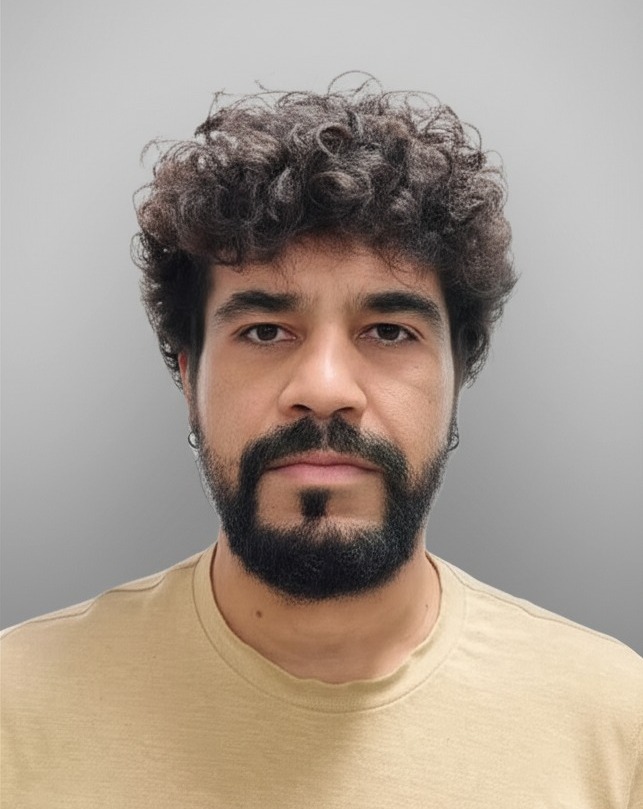}}]{Eduardo Ferreira da Silva}
Eduardo Ferreira is a Ph.D. student in Computer Science at the Federal University of Bahia. He holds an M.Sc. in Computer Science from the Federal University of Santa Maria. His research interests include recommender systems, machine learning, deep learning, and natural language processing. Contact him at eduardoferreira@ufba.br.
\end{IEEEbiography}

\begin{IEEEbiography}[{\includegraphics[width=1in,height=1.25in,clip,keepaspectratio]{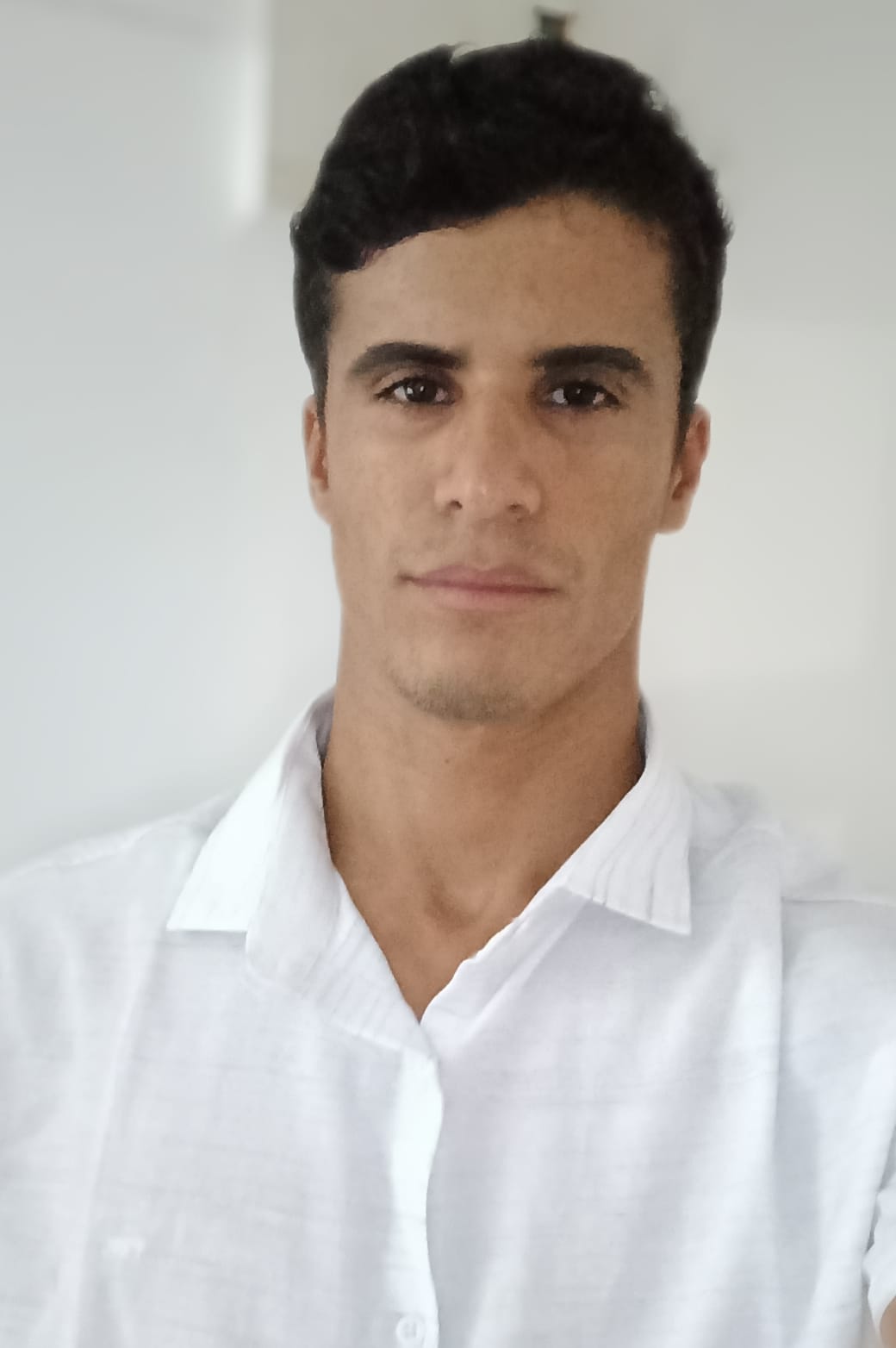}}]{Joel Machado Pires}
Joel Machado Pires is an MSc student at the Federal University of Bahia (UFBA), where he participates in RECSYS Lab. He has experience in  Machine Learning and Software Development, with applications in computer vision, forecasting, and recommendation systems. His current research field is reliability in recommender systems.
\end{IEEEbiography}

\begin{IEEEbiography}[{\includegraphics[width=1in,height=1.25in,clip,keepaspectratio]{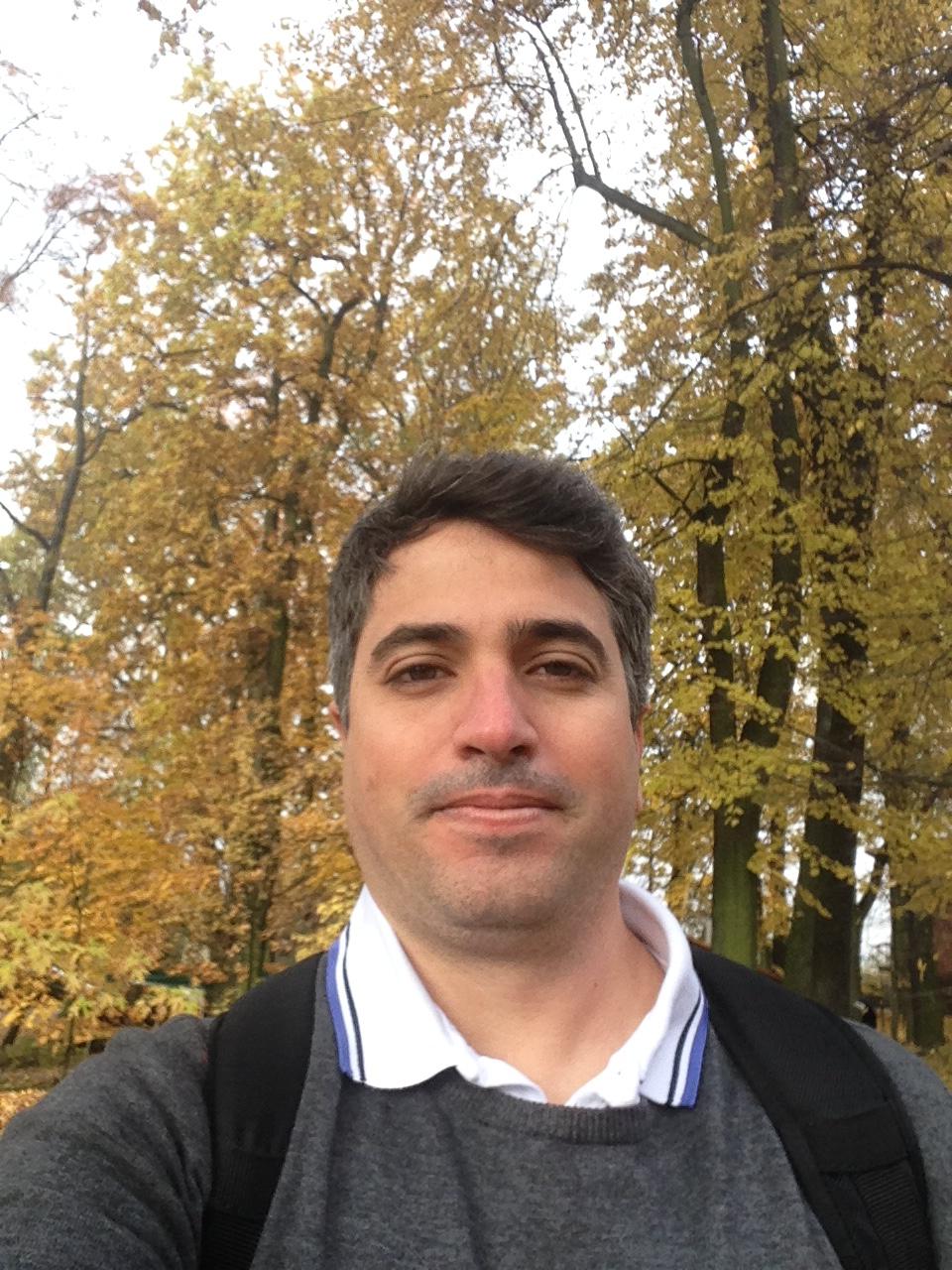}}]{Frederico Araújo Durão}
Frederico Araújo Durão is an associate professor at the Federal University of Bahia (UFBA), where he leads the RecSys Lab. His main research areas include: Recommender Systems, Web Engineering, and Semantic Web.
\end{IEEEbiography}

\vfill

\end{document}